\documentstyle[epsf]{mn}
\newif\ifAMStwofonts
\AMStwofontstrue

\ifoldfss
  \ifCUPmtlplainloaded \else
    \NewTextAlphabet{textbfit} {cmbxti10} {}
    \NewTextAlphabet{textbfss} {cmssbx10} {}
    \NewMathAlphabet{mathbfit} {cmbxti10} {} % for math mode
    \NewMathAlphabet{mathbfss} {cmssbx10} {} %  "   "    "
  \fi
  \ifAMStwofonts
    \ifCUPmtlplainloaded \else
      \NewSymbolFont{upmath} {eurm10}
      \NewSymbolFont{AMSa} {msam10}
      \NewMathSymbol{\upi}     {0}{upmath}{19}
      \NewMathSymbol{\umu}     {0}{upmath}{16}
      \NewMathSymbol{\upartial}{0}{upmath}{40}
      \NewMathSymbol{\leqslant}{3}{AMSa}{36}
      \NewMathSymbol{\geqslant}{3}{AMSa}{3E}

    \fi
  \fi
\fi % End of OFSS

\ifnfssone
  \newmathalphabet{\mathit}
  \addtoversion{normal}{\mathit}{cmr}{m}{it}
  \addtoversion{bold}{\mathit}{cmr}{bx}{it}
  \newmathalphabet{\mathbfit} % math mode version of \textbfit{..}
  \addtoversion{normal}{\mathbfit}{cmr}{bx}{it}
  \addtoversion{bold}{\mathbfit}{cmr}{bx}{it}
  \newmathalphabet{\mathbfss} % math mode version of \textbfss{..}
  \addtoversion{normal}{\mathbfss}{cmss}{bx}{n}
  \addtoversion{bold}{\mathbfss}{cmss}{bx}{n}
  \ifAMStwofonts
    \ifCUPmtlplainloaded \else
      \UseAMStwoboldmath
      \makeatletter
      \new@mathgroup\upmath@group
      \define@mathgroup\mv@normal\upmath@group{eur}{m}{n}
      \define@mathgroup\mv@bold\upmath@group{eur}{b}{n}
      \edef\UPM{\hexnumber\upmath@group}
      \new@mathgroup\amsa@group
      \define@mathgroup\mv@normal\amsa@group{msa}{m}{n}
      \define@mathgroup\mv@bold\amsa@group{msa}{m}{n}
      \edef\AMSa{\hexnumber\amsa@group}
      \makeatother
      \mathchardef\upi="0\UPM19
      \mathchardef\umu="0\UPM16
      \mathchardef\upartial="0\UPM40
      \mathchardef\leqslant="3\AMSa36
      \mathchardef\geqslant="3\AMSa3E
    \fi
  \fi
\fi % End of NFSS release 1

\ifnfsstwo
  \DeclareMathAlphabet{\mathbfit}{OT1}{cmr}{bx}{it}
  \SetMathAlphabet\mathbfit{bold}{OT1}{cmr}{bx}{it}
  \DeclareMathAlphabet{\mathbfss}{OT1}{cmss}{bx}{n}
  \SetMathAlphabet\mathbfss{bold}{OT1}{cmss}{bx}{n}
  \ifAMStwofonts
    \ifCUPmtlplainloaded \else
      \DeclareSymbolFont{UPM}{U}{eur}{m}{n}
      \SetSymbolFont{UPM}{bold}{U}{eur}{b}{n}
      \DeclareSymbolFont{AMSa}{U}{msa}{m}{n}
      \DeclareMathSymbol{\upi}{0}{UPM}{"19}
      \DeclareMathSymbol{\umu}{0}{UPM}{"16}
      \DeclareMathSymbol{\upartial}{0}{UPM}{"40}
      \DeclareMathSymbol{\leqslant}{3}{AMSa}{"36}
      \DeclareMathSymbol{\geqslant}{3}{AMSa}{"3E}
    \fi
  \fi
\fi % End of NFSS release 2

\ifCUPmtlplainloaded \else
  \ifAMStwofonts \else % If no AMS fonts
    \def\upi{\pi}
    \def\umu{\mu}
    \def\upartial{\partial}
  \fi
\fi
\title{Variograms of the Cosmic Microwave Background Temperature Fluctuations: Confirmation of Deviations from Statistical Isotropy}
\author[] {
L. Cay\'on$^{1}$\\
1. Department of Physics. Purdue University. 525 Northwestern Avenue, West Lafayette, IN 47907-2036, USA\\}

\date{\today}

\pagerange{\pageref{firstpage}--\pageref{lastpage}}
\pubyear{2001}

\begin{document}

\maketitle

\label{firstpage}

\begin{abstract}

\noindent  The Standard Inflationary model predicts an isotropic distribution of the Cosmic Microwave Background temperature fluctuations. Detection of deviations from statistical isotropy would call for a revision of the physics of the early universe. This paper introduces the variogram as a powerful tool to detect and characterize deviations from statistical isotropy in Cosmic Microwave Background maps. Application to the Wilkinson Microwave Anisotropy Probe data clearly shows differences between the northern and the southern hemispheres. The sill and range of the northern hemisphere's variogram are lower than those of the southern hemisphere. Moreover the variogram for the northern hemisphere lies outside the $99\%$ c.l. for scales above ten degrees. Differences between the northern and southern hemispheres in the functional dependence of the variogram with the scale can be used as a validation bench mark for proposed anisotropic cosmological models.

\end{abstract}

%\begin{keywords}
%cosmology: CMB -- data analysis
%\end{keywords}

\section{Introduction}

The most recent observations of the Cosmic Microwave Background (CMB) in combination with Large Scale Structure and High Redshift Supernova observations point toward a model of the universe with an energy density dominated by the Cosmological Constant, the so called concordance model (Dunkley et al. 2009). This model relies on the assumption of an early period of Inflation (Baumann \& Peiris 2009). The Standard Inflationary model predicts a isotropic Gaussian statistical distribution of the CMB fluctuations. Since the release of the first year of data of the Wilkinson Microwave Anisotropy Probe (WMAP) satellite there has been a substantial amount of work aimed at testing these two predictions (see Cay\'on 2007 for a list of works in real, spherical harmonic and wavelet space). In particular deviations from isotropy have been confirmed in the three and five years of WMAP data (Eriksen et al. 2007, Hansen et al. 2009, Rath et al. 2007, Vielva et al. 2007, Monteser\'\i n et al. 2008, Samal et al. 2008, Hoftuft et al. 2009). All these deviations should be seriously considered as they might be pointing toward the need to revise the physics behind the processes that take place in the early universe.

This paper introduces the variogram as a statistical tool to study the isotropy of CMB fluctuations. The variogram is a well known statistical method used in the analysis of spatially distributed data Schabenger \& Gotway 2005. It represents a direct method to estimate the correlation length of a sample. Its graphical representation provides information about the spatial continuity or roughness of a 2D image. In particular this method has been extensively applied in geostatistical analyses to quantify spatial variability (Clark 1979, Clark \& Harper 2000, Gringarten \& Deutsch 2001). Sample variograms in different directions are a conventional practice for characterizing the correlation structure of geological images and for checking for their isotropy. Variograms of CMB maps are shown to be easy to calculate and to interpret. The scale dependence of the variogram is shown to differ between the northern and southern hemispheres for the WMAP data revealing it as a powerful tool to validate non-isotropic cosmological models.

This paper is organized as follows. The variogram is defined in Section 2. Application of variograms to the study of CMB maps, in particular those provided by the WMAP mission, is addressed in Section 3. Statistical analysis of WMAP in comparison to Monte Carlo simulations is presented in Section 4. Section 5 is dedicated to discussion and conclusions.

\setcounter{figure}{0}
\begin{figure*}
 \epsfxsize=95mm
 \epsffile{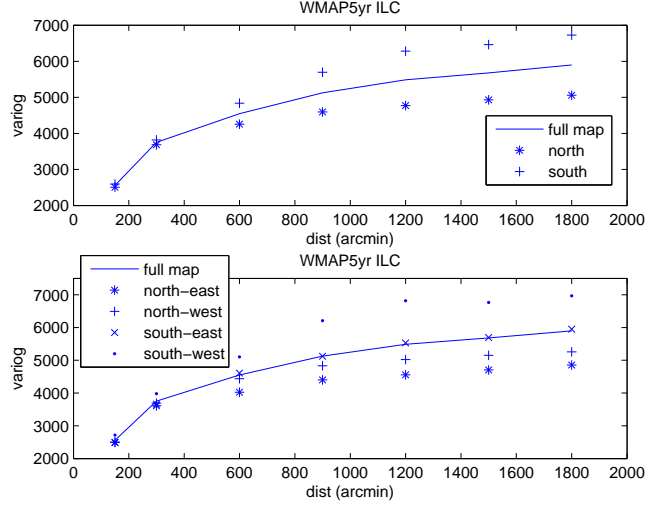}
 \caption{Variograms (in units of $\mu K^2$) of the ILC map. The top panel shows the variogram calculated over the entire map together with the variograms obtained accounting for pixels only in the northern (galactic latitude between $0$ and $\pi /2$ rads) or southern hemispheres (galactic latitude between $\pi /2$ and $\pi$ rads). The bottom panel shows the variogram calculated over the entire map together with the variograms corresponding to the northeast (in the northern hemisphere and with galactic longitude between $0$ and $\pi$ rads), northwest (in the northern hemisphere and with galactic longitude between $\pi$ and $2\pi$ rads), southeast (in the southern hemisphere and with galactic longitude between $0$ and $\pi$ rads) and southwest (in the southern hemisphere and with galactic longitude between $\pi$ and $2\pi$ rads).}
 \label{f1}
\end{figure*}

\setcounter{figure}{1}
\begin{figure*}
 \epsfxsize=95mm
 \epsffile{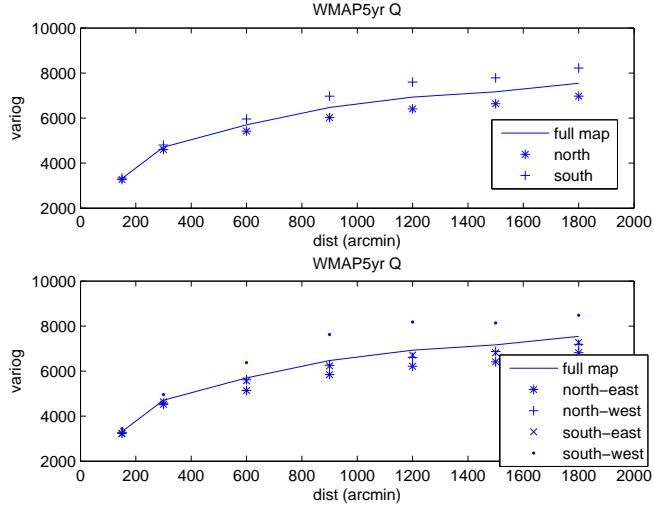}
 \caption{Variograms (in units of $\mu K^2$) of the Q map. The top panel shows the variogram calculated over the entire map together with the variograms obtained accounting for pixels only in the northern (galactic latitude between $0$ and $\pi /2$ rads) or southern hemispheres (galactic latitude between $\pi /2$ and $\pi$ rads). The bottom panel shows the variogram calculated over the entire map together with the variograms corresponding to the northeast (in the northern hemisphere and with galactic longitude between $0$ and $\pi$ rads), northwest (in the northern hemisphere and with galactic longitude between $\pi$ and $2\pi$ rads), southeast (in the southern hemisphere and with galactic longitude between $0$ and $\pi$ rads) and southwest (in the southern hemisphere and with galactic longitude between $\pi$ and $2\pi$ rads).}
 \label{f1}
\end{figure*}

\section{Variogram: Definition and Interpretation}

Given a stationary random field $Z$, the variogram for all possible locations $\vec u$ is defined as 
\begin{equation}
2\gamma(\vec h)=E[(Z(\vec u)-Z(\vec u+\vec h))^2].
\end{equation}
Where $Z(\vec u)$ represents the value of the random field at position $\vec u$ and $\vec h$ is the lag distance. $\gamma$ is usually called the semi-variogram. The variogram therefore measures changes in the field at different scales (lag distances). In particular
\begin{equation}
2\gamma(\vec h)=2[C(0)-C(\vec h)],
\end{equation} 
where $C$ is the correlation function. The sill of the variogram corresponds to the value of the variogram for which $C(\vec h^*)=0$. The lag distance $\vec h^*$ is called the range. A variogram that reaches a constant value for distances larger than the range reflects lack of correlation at those scales. Variograms represent a robust, fast and easy to implement method to determine the correlation length of a given sample without the need to compute the full correlation function.

A key question in geophysics and, as discussed in the introduction, in cosmology/astrophysics, is to check for whether the correlation structure in an image is isotropic or not. The correlation structure in an image would be isotropic if it depends only on the distance between two sites and not on their relative orientation. A common practice in geophysics is to determine the variogram along different directions (usually horizontal and vertical). Differences in the sill and range of variograms in different directions are indicators of geologic variability. We discuss in the next section how to implement this methodology in the study of statistical properties of CMB maps. 

\setcounter{figure}{2}
\begin{figure*}
 \epsfxsize=95mm
 \epsffile{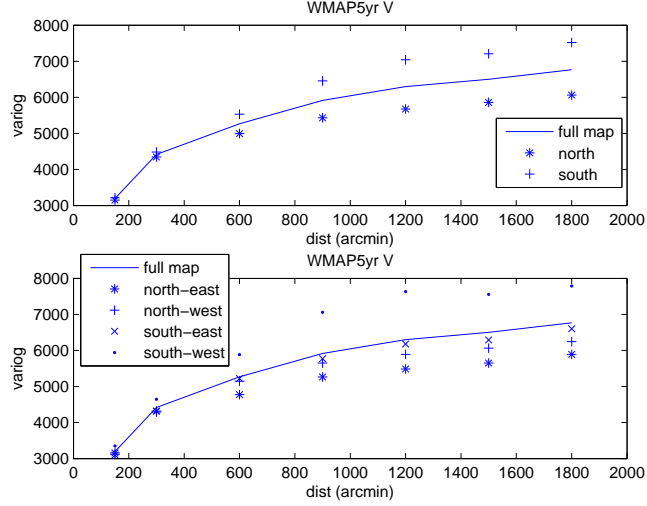}
 \caption{Variograms (in units of $\mu K^2$) of the V map. The top panel shows the variogram calculated over the entire map together with the variograms obtained accounting for pixels only in the northern (galactic latitude between $0$ and $\pi /2$ rads) or southern hemispheres (galactic latitude between $\pi /2$ and $\pi$ rads). The bottom panel shows the variogram calculated over the entire map together with the variograms corresponding to the northeast (in the northern hemisphere and with galactic longitude between $0$ and $\pi$ rads), northwest (in the northern hemisphere and with galactic longitude between $\pi$ and $2\pi$ rads), southeast (in the southern hemisphere and with galactic longitude between $0$ and $\pi$ rads) and southwest (in the southern hemisphere and with galactic longitude between $\pi$ and $2\pi$ rads).}
 \label{f1}
\end{figure*}

\setcounter{figure}{3}
\begin{figure*}
 \epsfxsize=95mm
 \epsffile{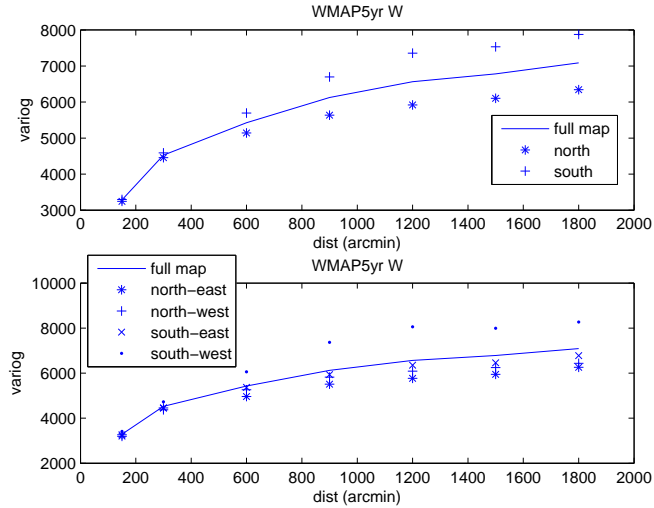}
 \caption{Variograms (in units of $\mu K^2$) of the W map. The top panel shows the variogram calculated over the entire map together with the variograms obtained accounting for pixels only in the northern (galactic latitude between $0$ and $\pi /2$ rads) or southern hemispheres (galactic latitude between $\pi /2$ and $\pi$ rads). The bottom panel shows the variogram calculated over the entire map together with the variograms corresponding to the northeast (in the northern hemisphere and with galactic longitude between $0$ and $\pi$ rads), northwest (in the northern hemisphere and with galactic longitude between $\pi$ and $2\pi$ rads), southeast (in the southern hemisphere and with galactic longitude between $0$ and $\pi$ rads) and southwest (in the southern hemisphere and with galactic longitude between $\pi$ and $2\pi$ rads).}
 \label{f1}
\end{figure*}

\section[]{Analysis of CMB maps based on variograms}

Observations of the CMB (maps) are distributed over the surface of the sphere. Common applications of the variogram methodology have been done on 2D flat images. Here we extend this statistical method to the analysis on the sphere and present the application to WMAP maps (Hinshaw et al. 2009, ApJS, Hinshaw et al. 2007).

The value of the variogram for the CMB temperature fluctuations $\Delta T$ at a lag distance $h$ is numerically calculated as 
\begin{equation}
2\gamma(h)=(1/N_p)\sum_{\vec u}[(1/N(\vec u,h))\sum_{N(\vec u, h)}[\Delta T(\vec u)-\Delta T(\vec u+h)]^2,
\end{equation}
where $N_p$ is the total number of central pixels (at position $\vec u$). $N(\vec u,h)$ indicates the number of pixels at a distance (angular distance) $h$ from the central pixel (at $\vec u$). The expression above is usually referred to as the omni-directional variogram.

\setcounter{figure}{4}
\begin{figure*}
 \epsfxsize=95mm
 \epsffile{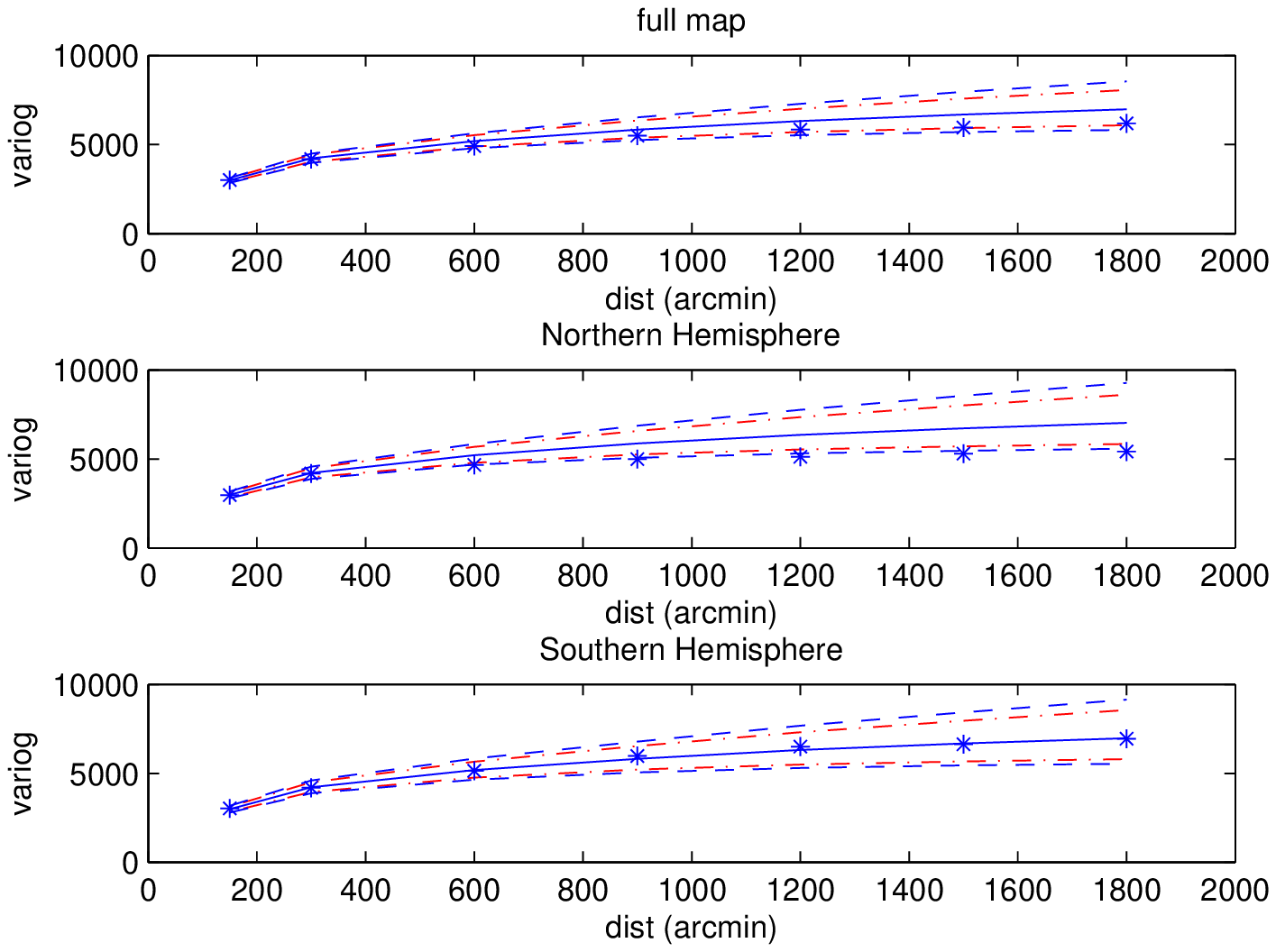}
 \caption{Variograms (in units of $\mu K^2$) of the full map and the northern and southern hemispheres. Stars correspond to the WMAP data. The solid line represents the average variogram obtained from 10000 simulations. Dotted-dashed and dashed lines correspond to $95\%$ and $99\%$ confidence limits respectively.}
 \label{f1}
\end{figure*}

\setcounter{figure}{5}
\begin{figure*}
 \epsfxsize=95mm
 \epsffile{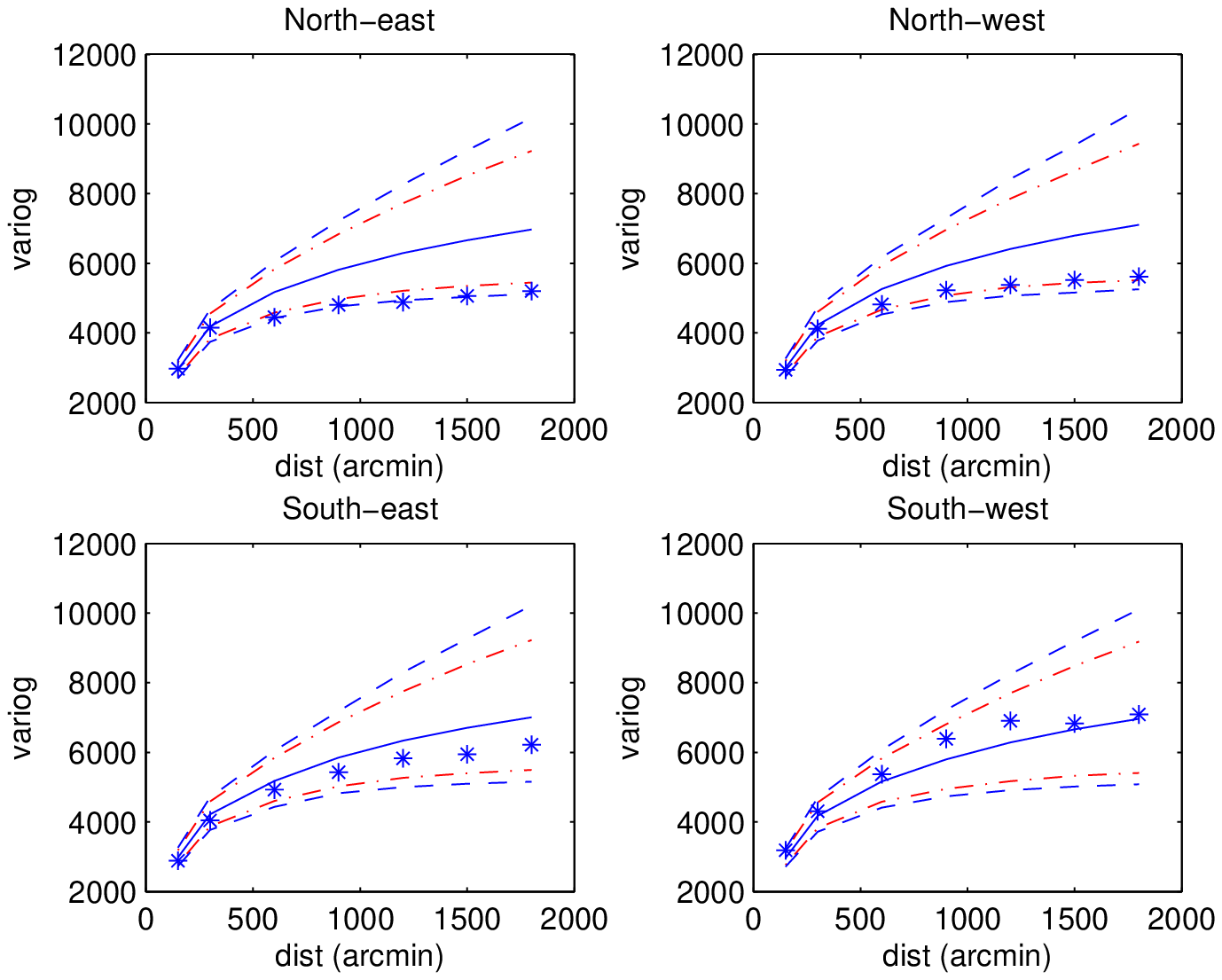}
 \caption{Variograms (in units of $\mu K^2$) of the northeast, northwest, southeast and southwest regions on the sphere. Stars correspond to the WMAP data. The solid line represents the average variogram obtained from 10000 simulations. Dotted-dashed and dashed lines correspond to $95\%$ and $99\%$ confidence limits respectively.}
 \label{f1}
\end{figure*}

Variograms for WMAP \footnote[1]{Data has been downloaded from http://lambda.gsfc.nasa.gov/} five year data are presented in Figures 1-4. Solid lines indicate the value of the omni-directional variograms. Figure 1 corresponds to the Internal Linear Combination map. Variograms for the individual frequency maps Q,V and W are presented in figures 2,3 and 4 respectively. In order to remove Galactic and point source emission, we applied the 
KQ85 mask provided by the WMAP team. Unless otherwise stated, only pixels outside the mask enter the computation of the variogram (either as central pixels or as pixels at a certain angular distance from the central one.) The maps were originally downloaded in the HEALpix\footnote[2]{HEALPix http://healpix.jpl.nasa.gov/} nested format at resolution $nside=512$ (total number
of pixels being $12\times nside^2$). Before the variograms were calculated 
we downgraded the maps to resolution $nside=32$. The minimum lag distance is constrained by the pixel size in that resolution. We calculate the variogram at lag distances $h$ between $150$ arcmins and $1800$ arcmins. In calculating the pixels that are at these distances from the central pixel we include all pixels in a ring with a lag tolerance of $30$ arcmins ($\pm 30$ arcmins around the fixed lag distance.) As one can see the sill of the variograms is being approached at scales above $\sim 1200$ arcmins. There is no difference in the scale dependence of the variogram among all the maps considered.

The standard method used in geostatistics to look for deviations from isotropy is to calculate the variogram in different directions. In applying this method to the analysis of CMB images on the surface of a sphere we have slightly modified the standard approach for considering different directions. In order to look for deviations from isotropy we have calculated variograms in different regions of the sphere. In galactic coordinates, the map was divided into the northern and southern hemispheres. One can notice two main differences between the two hemispheres. (1) The sill of the variogram corresponding to the northern hemisphere is reached at a slight lower range than that of the southern hemisphere and that of the entire map. (2) The value of the sill is lower in the north than in the south, pointing to a lower value of the variance. A similar trend is observed when the map is divided into four regions corresponding to the northeast, northwest, southeast and southwest (see the definition in figure captions for Figures 1-4). The range and sill appear to be the smallest for the northeast region. These results indicate a difference in the behavior of the correlation function in between the northern and southern hemispheres. For the northern hemisphere the correlation function drops to nearly zero at scales above $\sim 900$ and the variance is lower than that for the southern hemisphere.

\section{Statistical Analysis}

In order to determine the statistical significance of the observed anisotropy in the correlation function, we have determined the distribution of the variogram for full maps and for different regions of the sphere. The statistical distribution was built based on $10000$ simulations following the same procedure as in Cruz et al. 2007 and Vielva et al. 2004. The data and the simulations correspond to a weighted linear combination of WMAP Q, V and W bands (notice that the data maps used in this analysis are not those corresponding to the ILC provided by the WMAP team). These simulations included the number of observations and receiver noise dispersion corresponding to the three years of observations of WMAP (as provided in the LAMBDA web site). At the resolution ($nside=32$) at which the analysis was performed there should be no difference between three or five years of data (one can compare the variograms for WMAP in figures 1 and 5).

Variograms for the northern and southern hemispheres (galactic coordinates) are presented in Figure 5.  
The mean, $95\%$ and $99\%$ c.l. for the simulations are shown as solid, dotted-dashed and dashed lines respectively. Results for WMAP are denoted by stars. 
As one can see, the variogram for the southern hemisphere perfectly follows the behavior of the simulations. This is not the case for the northern hemisphere. The variogram for the northern hemisphere deviates from the expected value by more than $99\%$ at scales above $600$ arcmins. This is a consequence of lower range (lack of correlation above that scale) and sill for WMAP in comparison with the simulations. 

Variograms for the four regions, northeast, northwest, southeast, southwest, are shown in Figure 6. The results indicate that variograms for both northern regions are at the tails of the distribution. The range and sill for the northeast are smaller than those for any of the other three regions indicating that the north/south asymmetry could be originating from that region.

\section{Discussion and Conclusions}

The Standard Inflationary model predicts a universe that is statistically homogeneous and isotropic. There have been a number of works on detections of deviations from isotropy based on analyses of the WMAP data (see references in the introduction). In this paper we have introduced the variogram as a tool for detecting deviations from isotropy in CMB maps. Variograms have been extensively used in geological applications in order to characterize the soil in different directions. The variogram provides an alternative, easy to apply and easy to interpret method, to characterize the correlation function of the CMB temperature fluctuations. Under the assumption of isotropy there should be no difference in the behavior of the correlation function with direction.

Previous studies of the correlation function of the WMAP data (Copi et al. 2009) have shown a lack of correlation at scales above $\sim 60\deg$. This agrees with the functional dependence of the variogram observed in figures 1-4 and top pannel of figure 5. The variogram of the WMAP data starts converging toward a constant value at scales $\sim 30\deg$ with an expected value of the range at the scales indicated by the work of Copi et al. 2009. The functional dependence of the variogram differs from that expected in the corcondance model as seen in figure 5. The WMAP data presents shorter range and lower variance than the best fit cosmological model.  

The variograms of the WMAP data clearly show differences between the northern and southern hemisphere (see upper panels in figures 1-4). The northern hemisphere has lower sill and range than the southern hemisphere. This result points in the same direction as previous claims of lack of large and intermediate scale fluctuations in the northern hemisphere (Eriksen et al. 2004, Hansen et al. 2004, Eriksen et al. 2005, Eriksen et al. 2007, Monteser\'\i n et al. 2008, Hansen et al. 2009). Results presented in the lower panels of Figures 1-4 are also indicative of a larger lack of correlation in the northeast region in comparison with the other three regions of the sphere. The observed trend in the variogram is the same in the ILC map and the three band maps Q,V and W indicating no frequency dependence. 

Assessment of the statistical significance of the findings discussed above was done based on the variogram distribution function obtained from 10000 simulations. Simulations were performed assuming the concordance model and taking into account observational constraints from WMAP. The variogram of the northern hemisphere is outside the $99\%$ c.l. for scales above ten degrees (see Figure 5). The scale dependence of the variogram is clearly different between the northern hemisphere and the southern hemisphere as well as between the northern hemisphere and the simulations in both hemispheres. The variogram for the WMAP data in the northern hemisphere rises up to a scale of 300 arcmins, leveling off right above that scale. Variograms of the simulations and of the WMAP data in the southern hemisphere continue rising above 300 arcmins. The sill is directly related to the variance and these results therefore confirm a lower variance in the northern hemisphere, mostly concentrated in the northeast region.

This paper contributes to increase the evidence against statistical isotropy in the WMAP data. Results from the variogram are directly obtained in real space and easy to interpret. The difference in functional shape (scale dependence) of the variogram between the northern and southern hemispheres can be a validation bench mark for the predictions of non-isotropic cosmological models like some those already considered in the literature (Jaffe et al. 2005, 2006, Gordon et al. 2005, Ackerman et al. 2007). Consideration of some of these models is left for a future paper.

\section*{Acknowledgements}

We acknowledge the use of the Legacy Archive for Microwave Background Data Analysis (LAMBDA). Support for LAMBDA is provided by the NASA Office of Space
Science. 
Some of the results in this paper have been derived using the HEALPix (G\'orski, Hivon, and Wandelt 1999) package.


\begin{thebibliography}{}

\bibitem{} Ackerman, L., Carroll, S.M. \& Wise, M.B. 2007, Phys.Rev.D, 75, 083502

\bibitem{} Baumann, D. \& Peiris, H.V. 2009, Advanced Science Letters, 2, 105


\bibitem{} Cay\'on, L., Jin, J. \& Treaster, A. 2005, 362, 826


\bibitem{} Cay\'on, L. 2007, Statistical Challenges in Modern Astronomy IV, ASP Conference Series, Ed. Babu, G.J. \& Feigelson, E.D., 371, 18 

\bibitem{} Clark, I. 1979, {\it Practical Geostatistics}, Applied Science Publishers

\bibitem{} Clark, I., Harper, W.V. 2000, {\it Practical Geostatistics 2000}, Ecosse North America, LLC 

\bibitem{} Copi, C.J., Huterer, D., Schwarz, D.J. \& Starkman, G.D. 2009, MNRAS, 399, 295

\bibitem{} Cruz, M., Cay\'on, L., Mart\'\i nez-Gonz\'alez, E., Vielva, P. \& Jin, J. 2007, ApJ, 655, 11


\bibitem{} Dunkley, J. et al. 2009, ApJS, 180, 306

\bibitem{} Eriksen, H.K., Hansen, F.K., Banday, A.J., Gorski, K.M. \& Lilje, P.B. 2004, ApJ, 605, 14

\bibitem{} Eriksen, H.K., Banday, A.J., Gorski, K.M. \& Lilje, P.B. 2005, ApJ, 622, 58

\bibitem{} Eriksen, H.K., Banday, A.J., Gorski, K.M., Hansen, F.K. \& Lilje, P.B. 2007, ApJ, 660, L81


\bibitem{} Gordon, C., Hu, W., Huterer, D. \& Crawford, T. 2005, Phys.Rev.D., 72, 103002


\bibitem{} Gringarten, E. \& Deustch, C.V. 2001, {\it Mathematical Geology}, 33, 507


\bibitem{} Hansen, F.K., Banday, A.J. \& Gorski, K.M. 2004, MNRAS, 354, 641

\bibitem{} Hansen, F.K., Banday, A.J., Gorski, K.M., Eriksen, H.K. \& Lilje, P.B. 2009, ApJ, 704, 1448


\bibitem{} Hinshaw, G. et al. 2007, ApJSS, 170, 288

\bibitem{} Hinshaw, G. et al. 2009, ApJSS, 180, 225

\bibitem{} Hoftuft, J., Eriksen, H.K., Banday, A.J., Gorski, K.M., Hansen, F.K.\& Lilje, P.B. 2009, ApJ, 699, 985


\bibitem{} Jaffe, T.R., Banday, A.J., Eriksen, H.K., Gorski, K.M. \& Hansen, F.K. 2005, ApJ, 629, L1

\bibitem{} Jaffe, T.R., Banday, A.J., Eriksen, H.K., Gorski, K.M. \& Hansen, F.K. 2006, A\&A, 460, 393

\bibitem{} Monteser\'\i n, C., Barreiro, R.B., Vielva, P., Mart\'\i nez-Gonz\'alez, E., Hobson, M.P. \& Lasenby, A.N. 2008, MNRAS, 387, 209

 
\bibitem{} Rath, C., Schuecker, P., Banday, A.J. 2007, MNRAS, 380, 466

\bibitem{} Samal, P.K., Saha, R., Jain, P. \& Ralston, J.P. 2008, MNRAS, 385, 1718


\bibitem{} Schabenger, O., Gotway, C.A. 2005, {\it Statistical Methods for Spatial Data Analysis}, ed. Chapman \& Hall, 133

\bibitem{} Vielva, P., Mart\'\i nez-Gonz\'alez, E., Barreiro, R.B., Sanz, J.L. \& Cay\'on, L. 2004 ApJ, 609, 22

\bibitem{} Vielva, P., Wiaux, Y., Mart\'\i nez-Gonz\'alez, E., Vandergheynst, P. 2007, MNRAS, 381, 932

\end{thebibliography}
\end{document}